\RequirePackage{fix-cm}
\documentclass[twocolumn,epjc3]{svjour3}  
\smartqed  

\RequirePackage{graphicx}
\RequirePackage{dcolumn}
\RequirePackage{bm}
\RequirePackage{hyperref}
\RequirePackage{color} 
\RequirePackage{xcolor} 
\RequirePackage{color} 
\RequirePackage{graphics}
\RequirePackage{float}
\RequirePackage{bm}        
\RequirePackage{amssymb}   
\RequirePackage{slashed}   
\RequirePackage{amsmath}   
\RequirePackage{verbatim}  
\RequirePackage{subfig}
\RequirePackage{mathptmx}      
\RequirePackage{flushend}
\RequirePackage[numbers,sort&compress]{natbib}

\newcounter{YJC}

\journalname{Eur. Phys. J. C}

\begin{document}

\title{Using k-means assistant event selection strategy to study anomalous quartic gauge couplings at muon colliders}

\author{Shuai Zhang\thanksref{addr1,addr2}
        \and
        Ji-Chong Yang\thanksref{e1,addr1,addr2}
        \and
        Yu-Chen Guo\thanksref{addr1,addr2}
}

\thankstext{e1}{Corresponding author, e-mail: yangjichong@lnnu.edu.cn}

\institute{Department of Physics, Liaoning Normal University, Dalian 116029, China \label{addr1}
           \and
           Center for Theoretical and Experimental High Energy Physics, Liaoning Normal University, Dalian 116029, China \label{addr2}
}

\date{Received: date / Accepted: date}

\maketitle

\begin{abstract}
The search for new physics beyond the Standard Model is one of the central problems of current high energy physics interest. 
As the luminosities of current and near-future colliders continue to increase, the search for new physics has increased the requirements for processing large amounts of data.
Meanwhile, quantum computing which is rapidly evolving, has great potential to become a powerful tool to help search for new physics signals. 
Since the k-means algorithm is known to be able to be accelerated with the help of quantum computing, we investigate and propose an event selection strategy based on k-means algorithm to search for new physics signals. 
Taking the case of tri-photon processes at the muon colliders as an example, the event selection strategy is shown to be effective in helping to search for the signals of dimension-8 operators contributing to anomalous quartic gauge couplings.
Compared with traditional event selection strategy, the expected constraints are generally tighter.
\end{abstract}

\maketitle

\section{\label{sec1}Introduction}

While the Standard Model~(SM) has been very successful so far, there are still unanswered questions and the search for new physics~(NP) signals beyond the SM is at the forefront of high energy physics~(HEP)~\cite{johnellis}.
On the one hand, except for a few cases known or suspected to deviate from the SM~\cite{neutrinomass1,neutrinomass2,g2muon,rdstar2,wmass}, there is no clear signal giving guidance for possible NP models. 
Moreover, the experimental advances in the near decade ahead will be focused on the high luminosity frontier, which makes it important to search for NP efficiently.

A model-independent approach known as the SM effective field theory~(SMEFT) has attracted a lot of attention because of its ability to efficiently search for NP signals~\cite{weinberg,SMEFTReview1,SMEFTReview2,SMEFTReview3}. 
While the SMEFT is mainly applied to study the dimension-6 operators, dimension-8 operators are getting more and more attention recently~\cite{d81,vbs1,ssww,wastudy,wwstudy,zastudy}.
The study of the dimension-8 operators is necessary because of the convex geometry perspective to the operator space due to positivity bounds~\cite{positivity1,positivity2,positivity3}.
From a phenomenological point of view, there are many cases where dimension-6 operators are absent, but the dimension-8 operators show up~\cite{bi1,bi2,bi3,ntgc1,ntgc2,ntgc3,ntgc4,ntgc5,ntgc6,ntgc7}.
The experiments at the Large Hadron Collider~(LHC) also keep a close eye on dimension-8 operators~\cite{sswwexp1,sswwexp2,zaexp1,zaexp2,zaexp3,waexp1,zzexp1,zzexp2,wzexp1,wzexp2,wwexp1,wwexp2,wvzvexp,waexp2,zzexp3}.
However, for one generation of fermions, there are $895$ baryon number conserving dimension-8 operators~\cite{vbscan,d81}, and kinematic analysis needs to be done for each operator.
The efficiency decreases with the growth of the number of operators to be considered.

The use of machine learning~(ML) algorithms in HEP is recently developing rapidly~\cite{mlreview,ml1,ml2,ml3,ml4,ml5,ml6,ml8,ml10,ml11,wwww,wwwwunitary}.
To further improve efficiency, event selection strategies based on anomaly detection~(AD) ML algorithms are introduced to HEP community~\cite{ad,guassian,autoencoder1,autoencoder2,ml7,ml9,ml12}.
One advantage of using anomaly detection algorithms is that, usually no a priori knowledge of the NP models under study is required to find anomalous signals, i.e. the strategy to use AD is independent of the operators or NP models to be searched for.
Except for that, an AD algorithm may also serve as an aid to the traditional event selection strategies to improve the results.
Moreover, unlike searching for NP in a process that may turn out to be fruitless, the anomalies found by searching for anomalous signals are always worth attention.

Another promising improvement in efficiency comes from quantum computing~\cite{qc0,qc1,qc2}.
It has been shown that, the calculation of distances between two vectors can be accelerated by using the controlled swap gate which is known as swap test~\cite{swaptest,swaptest2}.
Meanwhile, the calculation of distances is at the core of the famous k-means algorithm~\cite{kmeans}, and the quantum accelerated k-means algorithms have been developed~\cite{qkmeans1,qkmeans2}.
The k-means algorithm can also been used for AD, however, whether the k-means AD algorithm~(KMAD) is useful in searching for NP is still not known.
In this paper, we focus on the feasibility of using KMAD in the search of NP.

As a test bed, we use KMAD to study the dimension-8 anomalous quartic gauge couplings~(aQGCs)~\cite{aqgcold,aqgcnew} at the muon colliders.
The aQGCs also draw a lot of attention at the LHC because they are closely related to the NP w.r.t. electroweak symmetry breaking~\cite{vbs1,vbs2,vbscan,positivity1,positivity2}.
Meanwhile, the muon colliders have received a lot of attention recently due to their abilities to reach both high energies and high luminosities, while keeping cleaner environment less affected by the QCD backgrounds~\cite{muoncollider1,muoncollider2,muoncollider3,muoncollider4,muoncollider5,muoncollider6,muoncollider7,muoncollider8,muoncollider9,muoncollider10}.
The tri-photon processes at the muon colliders are shown to be sensitive to the transverse operators contributing to the aQGCs~\cite{triphoton}.
Due to the narrow `EFT triangles' problem in the SMEFT~\cite{mo1,efttraingle2,efttraingle3}, i.e., if no new particles are found at very high energies and an EFT still applies, implying that the Wilson coefficients of the EFT have to be small or else the validity of the EFT would be violated, high luminance is necessary.
On the other hand, signatures of NP should be relatively easy to find, or already have been found, if they are not particularly exceptionally small. 
In this paper, we assume that the search for NP on muon colliders remains difficult, and therefore the search for NP remains a search for tiny signals in a large number of background events.
In this paper, we present an event selection strategy to search for the signals of aQGCs and set constraints on the operator coefficients with the help of KMAD.

The rest of this paper is organized as follows.
In Sec.~\ref{sec2}, the dimension-8 operators contributing to aQGCs and the tri-photon process are briefly reviewed.
The event selection strategy using KMAD is discussed in Sec.~\ref{sec3}.
The expected constraints on the operator coefficients are presented in Sec.~\ref{sec4}.
Sec.~\ref{sec5} summarizes our main conclusions.

\section{\label{sec2}The contribution of aQGCs to the tri-photon process}

\begin{table}[hbtp]
\centering
\begin{tabular}{c|c|c|c}
\hline
coefficient & constraint & coefficient & constraint \\
\hline
$f_{T_{0}}/\Lambda^4$ &[-0.12, 0.11]~\cite{wvzvexp}&$f_{T_{6}}/\Lambda^4$&[-0.4, 0.4]~\cite{waexp2} \\
$f_{T_{1}}/\Lambda^4$ &[-0.12, 0.13]~\cite{wvzvexp}&$f_{T_{7}}/\Lambda^4$&[-0.9, 0.9]~\cite{waexp2} \\
$f_{T_{2}}/\Lambda^4$ &[-0.28, 0.28]~\cite{wvzvexp}&$f_{T_{8}}/\Lambda^4$&[-0.43, 0.43]~\cite{zzexp3} \\
$f_{T_{5}}/\Lambda^4$ &[-0.5, 0.5]~\cite{waexp2}&$f_{T_{9}}/\Lambda^4$&[-0.92, 0.92]~\cite{zzexp3}\\
\hline
\end{tabular}
\caption{The constraints on the $O_{T_{i}}$ coefficients (${\rm TeV}^{-4}$) obtained at $95\%$ C.L at the LHC.}
\label{table:1}
\end{table}

The frequently used dimension-8 operators contributing to aQGCs can be classified as scalar/longitudinal operators $O_{S_i}$, mixed transverse and longitudinal operators $O_{M_i}$ and transverse operators $O_{T_i}$~\cite{vbscan}. At tree level, the tri-photon process at muon colliders can be affected by $O_{T_i}$ operators~\cite{aqgcold,aqgcnew},
\begin{equation}
\begin{split}
&O_{T,0}={\rm Tr}\left[\widehat{W}_{\mu\nu}\widehat{W}^{\mu\nu}\right]\times {\rm Tr}\left[\widehat{W}_{\alpha\beta}\widehat{W}^{\alpha\beta}\right],\\
&O_{T,1}={\rm Tr}\left[\widehat{W}_{\alpha\nu}\widehat{W}^{\mu\beta}\right]\times {\rm Tr}\left[\widehat{W}_{\mu\beta}\widehat{W}^{\alpha\nu}\right],\\
&O_{T,2}={\rm Tr}\left[\widehat{W}_{\alpha\mu}\widehat{W}^{\mu\beta}\right]\times {\rm Tr}\left[\widehat{W}_{\beta\nu}\widehat{W}^{\nu\alpha}\right],\\
&O_{T,5}={\rm Tr}\left[\widehat{W}_{\mu\nu}\widehat{W}^{\mu\nu}\right]\times B_{\alpha\beta}B^{\alpha\beta},\\
&O_{T,6}={\rm Tr}\left[\widehat{W}_{\alpha\nu}\widehat{W}^{\mu\beta}\right]\times B_{\mu\beta}B^{\alpha\nu},\\
&O_{T,7}={\rm Tr}\left[\widehat{W}_{\alpha\mu}\widehat{W}^{\mu\beta}\right]\times B_{\beta\nu}B^{\nu\alpha},\\
&O_{T,8}=B_{\mu\nu}B^{\mu\nu}\times B_{\alpha\beta}B^{\alpha\beta},\\
&O_{T,9}=B_{\alpha\mu}B^{\mu\beta}\times B_{\beta\nu}B^{\nu\alpha},\\
\end{split}
\label{eq.2.1}
\end{equation}
where $\widehat{W}\equiv \vec{\sigma}\cdot {\vec W}/2$ with $\sigma$ being the Pauli matrices and ${\vec W}=\{W^1,W^2,W^3\}$, $B_{\mu}$ and $W_{\mu}^i$ are $U(1)_{\rm Y}$ and $SU(2)_{\rm I}$ gauge fields, and $B_{\mu\nu}$ and $W_{\mu\nu}$ correspond to the field strength tensors.
The constraints on the coefficients obtained by the LHC are listed in Table~\ref{table:1}.

\begin{figure}[htbp]
\begin{center}
\includegraphics[width=0.98\hsize]{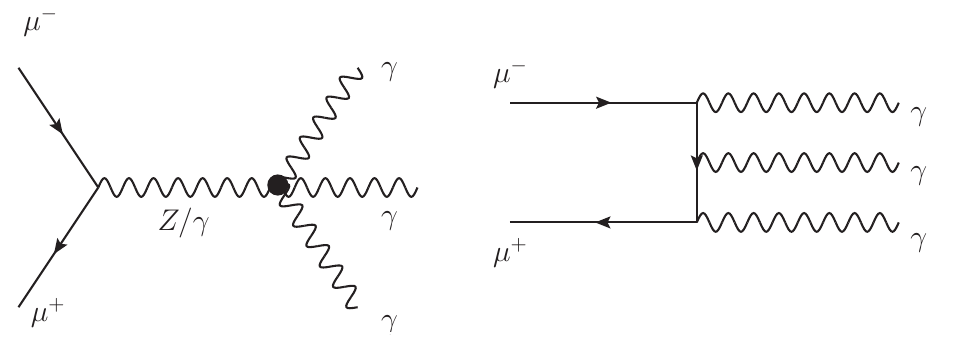}
\caption{\label{fig:Feyman}Feynman diagrams for the tri-photon processes $\mu^{+} \mu^{-} \rightarrow \gamma\gamma\gamma$ at the muon colliders. 
The diagrams induced by $O_{T_{i}}$ operators are shown in the left panel, and one of the diagrams in the SM is shown in the right
panel. 
In the SM, there are other five diagrams which can be obtained by permuting the photons in the final state.}
\end{center}
\end{figure}

The reason this set of operator basis is used is because this set has been used extensively at the LHC~\cite{sswwexp1,sswwexp2,zaexp1,zaexp2,zaexp3,waexp1,zzexp1,zzexp2,wzexp1,wzexp2,wwexp1,wwexp2,wvzvexp,waexp2,zzexp3} as well as in previous studies~\cite{wastudy,wwstudy,zastudy,wwwwunitary,ml6,ml7,triphoton,wwww,Dong:2023nir}, and therefore we can compare our results with previous studies.
The $O_{T_i}$ operators~(and linear combinations of $O_{T_i}$) are also in a complete basis of dimension-8 operators.
For example, $O_{T_0}=4O_{W^4}^{(1)}$, $O_{T_1}=4O_{W^4}^{(3)}$, $O_{T_2}=O_{W^4}^{(1)}+O_{W^4}^{(3)}+O_{W^4}^{(4)}$, $O_{T_5}=2O_{W^2B^2}^{(2)}$, $O_{T_6}=2O_{W^2B^2}^{(3)}$, $O_{T_7}=\left(O_{W^2B^2}^{(1)}+O_{W^2B^2}^{(3)}+O_{W^2B^2}^{(4)}\right)/2$, $O_{T_8}=4O_{B^4}^{(1)}$, and $O_{T_9}=\left(O_{B^4}^{(1)}/2\right)+\left(O_{B^4}^{(2)}/4\right)$ where $O_{W^4}^{(i)}$, $O_{W^2B^4}^{(i)}$ and $O_{B^4}^{(i)}$ are operators in the operator basis in Ref.~\cite{Murphy:2020rsh}, and they are also operators in the operator basis in Ref.~\cite{Li:2020gnx}. 

At tree level, there are only two Feynman diagrams induced by $O_{T_{i}}$ operators as shown on the left panel of Fig.~\ref{fig:Feyman}.
The SM background is shown on the right panel of Fig.~\ref{fig:Feyman}.
For the tri-photon process, the contribution of $O_{T_{1,6}}$ operators are exactly as same as $O_{T_{0,5}}$ operators, respectively.
In the following, we concentrate on $O_{T_{0,2,5,7,8,9}}$ operators.

\section{\label{sec3}K-means assistant event selection strategy}

\begin{figure}[htbp]
\begin{center}
\includegraphics[width=0.6\hsize]{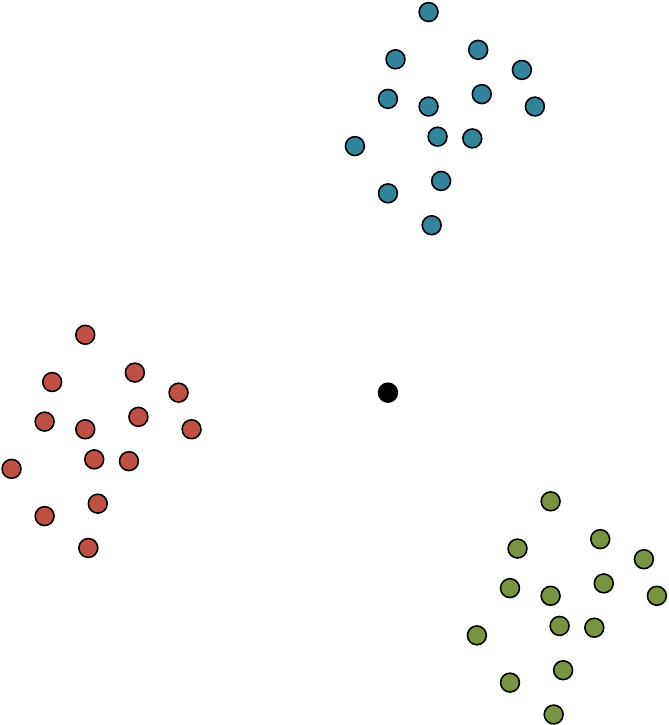}
\caption{\label{fig:illustrative}
A demonstration of the search for anomalous signals when the distribution of points is not Gaussian.
The black dot in the center is the anomalous point.
The blue, red and green points are normal points which can be divided into three clusters.
The anomalous point is far away from the centroids of all clusters.}
\end{center}
\end{figure}

\begin{figure*}[htbp]
\begin{center}
\includegraphics[width=0.4\hsize]{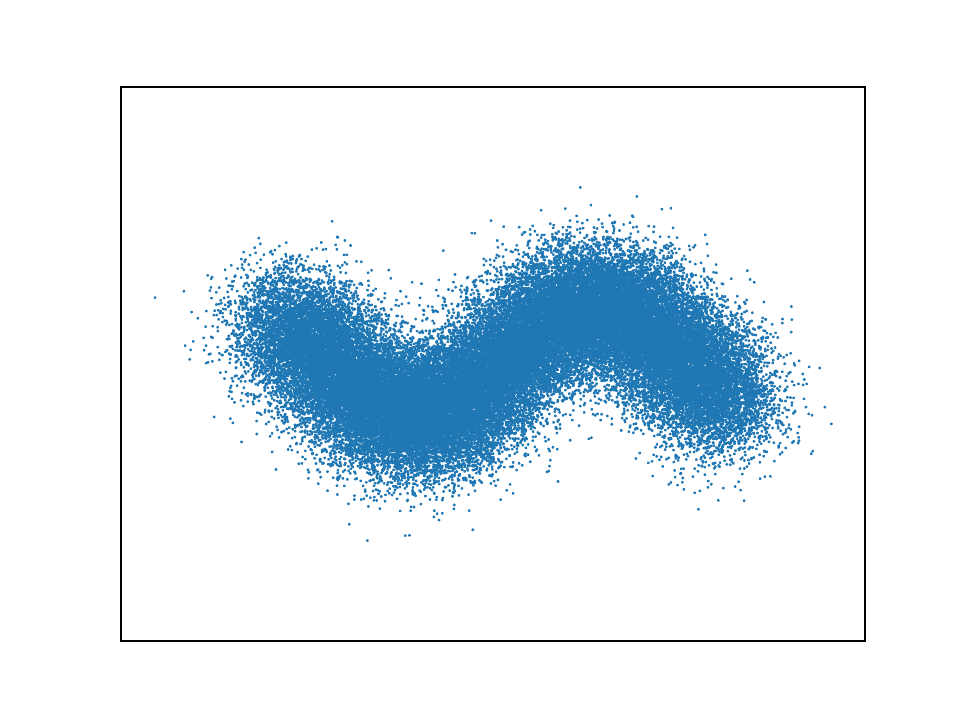}
\includegraphics[width=0.4\hsize]{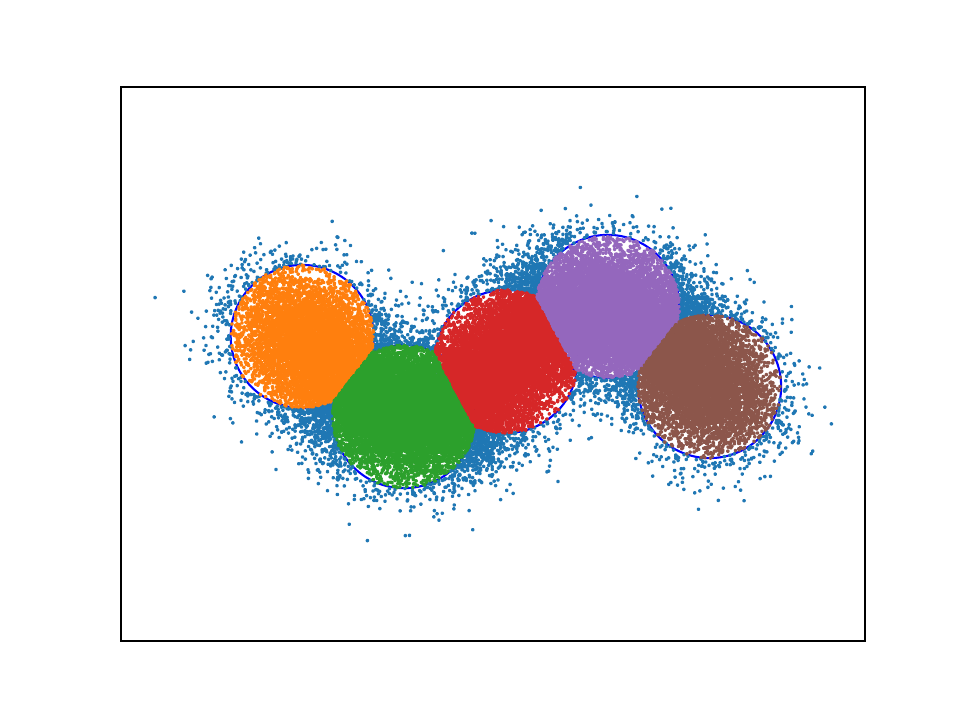}
\caption{\label{fig:illustrative2}
An example that the background events distribute arbitrarily~(the left panel).
One can still use a clustering algorithm to divide the background events into several classes~(the right panel).
The points farther from all centroids are more likely the anomalous events~(the blue dots in the right panel).}
\end{center}
\end{figure*}

Typically, the searching of NP at a collider with a high luminosity is to look for a small number of anomalies in the vast amount of data.
Meanwhile, AD algorithms are designed to search for those events which are `few and different'. 
Therefore, it can be expected that the AD algorithms are suitable to search for NP signals.

In the simplest case, one can assume that the elements in the dataset are Gaussian distributed.
The degree of anomaly of each event can be quantified by the distance of the point representing the event from the centroid of all points.
However, the picture of Gaussian distribution is often oversimplified for the case of searching for NP.
As an example, a more complicated case is shown in Fig.~\ref{fig:illustrative}.
The anomalous point is depicted as the black dot, which is approximately at the centroid of all points, therefore, the degree of anomaly can no longer be quantified by the distance of the point from the centroid of all points.
On the other hand, as shown in the Fig.~\ref{fig:illustrative}, the normal points can be divided into three clusters.
The anomalous point is far from the centroids of all clusters.
The exploitation of this feature is the key idea of KMAD.
By using KMAD, we use k-means to find the centroids of the clusters automatically.

\begin{figure*}[htbp]
\begin{center}
\includegraphics[width=0.98\textwidth]{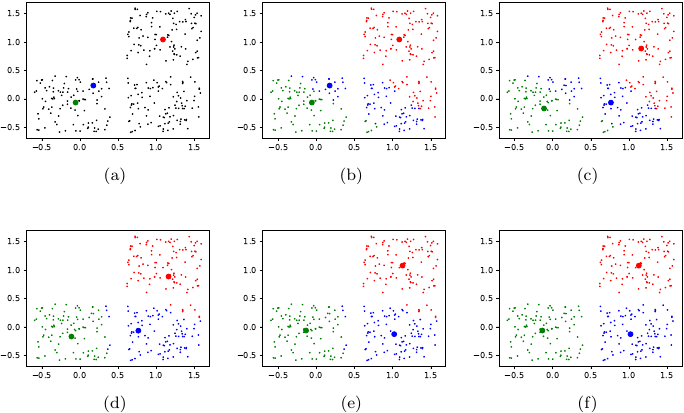}
\caption{\label{fig:illustrative3}
A sketch of how k-means cluster the data points with $k=3$.}
\end{center}
\end{figure*}  
A sketch of how k-means cluster the data points is plotted in Fig.~\ref{fig:illustrative3}.
At first, $k$~(in Fig.~\ref{fig:illustrative3}, $k=3$) centroids are chose randomly, which is depicted as the red, green and blue dots in Fig.~~\ref{fig:illustrative3}.~(a).
Then, the points are assigned in the same cluster as the closest centroid, which is shown in Fig.~~\ref{fig:illustrative3}.~(b).
And then the centroids are recalculated as the centers of the points with the same cluster assignments, which is depicted as the red, green and blue dots in Fig.~\ref{fig:illustrative3}~(c).
The steps in Figs.~\ref{fig:illustrative3}.~(b) and (c) are repeated until the assignment of the points do no longer change, as shown in Figs.~\ref{fig:illustrative3}.~(d), (e) and (f).
Calculating the distance from each point to each centroid in this procedure is the most computationally resource intensive, and calculating the distance can be accelerated with the swap test in quantum computing~\cite{qkmeans3}. 
Besides, swap test can directly calculate the distance from one vector to the centers of many vectors, and the cluster assignments can also be accelerated with a quantum computer~\cite{qkmeans1}. 

After clustering, we calculate the distances of the points to the centroids of all clusters.
Since the anomalous point is far from all centroids, the distance from a point to the nearest centroid can be used to quantify the degree of anomaly of the point.

This mechanism is suitable for even more general cases. 
Assume that the background events distribute arbitrarily as the one shown in the left panel of Fig.~\ref{fig:illustrative2}. 
If the classes obtained using the clustering algorithm can put together the shape of the distribution of background events, one can still use KMAD to improve the signal significance.
As shown in the right panel of Fig.~\ref{fig:illustrative2}, one can still choose the distance from a point to its nearest centroid as the anomaly score. Keeping only the points with large anomaly scores can significantly suppress the background events.
It can be expected that, the more clusters can sample an arbitrarily shaped distribution more accurately.
As will be shown, KMAD with more clusters works better in the case of searching for NP.

The essence behind the anomaly score mechanism is hypothesis testing. 
Still taking the simplest example of a Gaussian distribution, suppose that the SM predicts that the probability distribution of the occurrence of an event follows a Gaussian distribution, and suppose that these events can be described as points in a 2-dimensional space. The percentage of points that can be expected to be in a circle with a radius equal to twice of the standard deviation $2 \sigma$ is $95\%$, and the percentage of points in a circle with a radius equal to $3\sigma$ is $99.7\%$. 
Therefore, in experiments, when the number of points outside the $3\sigma$ circle exceeds $0.3\%$ of the total number of points, the distribution obtained by measurements must not be Gaussian, and the SM must need to be improved. 
Not only that, but since the points outside the $3\sigma$ circle account for $0.3\%$ of the total under the SM, the points outside the $3\sigma$ circle are more likely to be the signal of some unknown NP model than the points inside the circle. 
Thus the distance of an event from the center can be used as an anomaly score and become a quantitative criterion for portraying whether or not an event is more likely to be a NP event. 
KMAD, on the other hand, generalizes the above case of the Gaussian distribution to the case of more general distributions.  

\subsection{\label{sec3.1}Data preparation}

To prepare the dataset, we generate the events using Monte Carlo~(MC) simulation with the help of \verb"MadGraph5@NLO" toolkit~\cite{madgraph,feynrules,ufo}, including a muon collider-like detector simulation with \verb"Delphes"~\cite{delphes}.
To avoid infrared divergences, in this section we use the standard cut as default, the cuts relevant to infrared divergences are
\begin{equation}
\begin{split}
&p_{T,\gamma} > 10\;{\rm GeV},\;\; |\eta _{\gamma}| < 2.5, \;\; \Delta R_{\gamma\gamma} > 0.4,
\end{split}
\label{eq.standardcuts}
\end{equation}
where $p_{T,\gamma}$ and $\eta _{\gamma}$ are the transverse momentum and pseudo-rapidity for each photon, respectively, $\Delta R_{\gamma\gamma}=\sqrt{\Delta \phi ^2 + \Delta \eta ^2}$ where $\Delta \phi$ and $\Delta \eta$ are differences between the azimuth angles and pseudo-rapidities of two photons.
The events for signals are generated with one operator at a time.
In this section, the coefficients are chosen as the upper bound of Table~\ref{table:1}.
As will be introduced later, the coefficients as well as the operators to be searched for are actually irrelevant.

To build datasets for the KMAD, we require each event to have at least three photons.
The datasets are composed of the 4-momenta of the three hardest photons, i.e., we select three photons with the highest energy and rank them in descending order of energy, taking the components of the 4-momentum of each photon in the order, so that an event can be correspond to a 12-dimensional vector. 
In this section, we select 600000 events for the SM, and 200000 events are generated for the NP.
Since we only consider the difference between background features and signal features in this section, the interference is ignored.
The interference will be included in the next section.

In terms of the principle of the k-means algorithm, as long as the NP distribution is different from the background in the space of observables, it is possible to use the KMAD in this space of observables. 
The use of the KMAD and the space of observables can be combined at will. 
To show that KMAD is an automatic algorithm which has the advantage that one hardly needs to analyze the characteristics of the NP signal theoretically, we use momentum space straightforwardly.

It is important to emphasize that we only use the SM dataset in this step in training, and the NP datasets are used to shown the distribution of anomaly scores. 
Moreover, to remove the priori knowledge about this process as much as possible, after the datasets are built, the $12$ components are treated as just $12$ real numbers.
In fact the $12$ components are not independent of each other. 
One can make a step of analysis that does not depend on the physical content to remove redundancy. 
For example, one can use principal component analysis, or autoencode before proceeding to the next step. 
In this paper, we focus on the effect of KMAD and this step is not in our consideration for simplicity.
Again to avoid the use of prior knowledge, in this paper we simply use the Euclidean distance as the distance between two events.

\subsection{\label{sec3.2}Event selection strategy}

The number of clusters in the k-means algorithm is often denoted as the $k$, and the cluster assignment is often denoted as the k-value.
The following is a brief summary of the implementation of KMAD,
\begin{enumerate}
\item Specify a k-value for each point randomly. 
\item Calculate the cluster centroid as the center of the points with k-values as same as the cluster.
\item Calculate the distance from each point to the $k$ centroids~(all centroids) separately, and by comparing the distances, specify the k-value of the point as the k-value of the nearest centroid.
\item Repeat steps 2 and 3, until the k-value of each point is no longer changing.
\item Calculate the anomaly score of one point as the distance~(denoted as $d$) from the point to the centroid with the same k-value as this point.
\end{enumerate}

Denote the number of points whose k-value changed at the previous step as $n$, practically, to speed up the procedure, we stop when $n$ is less than $0.1\%$ of the total event number, i.e., when vast majority in cluster assignments no longer change.
There are cases where the classes corresponding to some k-values are empty.
In this case, we divide the largest cluster randomly and maintain the total number of clusters as $k$.
When the k-means process~(step 1 to 4) is finished, we can calculate how anomalous a point is, which is called anomaly score~($d$ in step 4).
Since the k-values given in the first step are random, and we stop before $n=0$, the results of the distances of the points to the centroids are not fixed.
To avoid the effect of these randomness, we repeated the above process $m$ times and make use of the average of the distances~(denoted as $\bar{d}$), i.e. the average anomaly score to distinguish the signal events from the background events.

If it is to find anomalous signals in a dataset, only the dataset needs to be provided without specifying the source of the dataset, and the physical content behind it. 
KMAD will automatically discover those anomalous events in the dataset that are different from the majority. 
The dataset can be the one obtained directly from experiment, without knowing the predictions from both the SM and NP.
Thus KMAD is an unsupervised learning algorithm. 
At the same time, KMAD can also be used as a supervised learning algorithm. 
In steps 1-4, the SM dataset obtained from MC can be used, and the process of obtaining centroids can be regarded as the training process. 
In step 5, the dataset from the experiment is used to find signals in the experimental dataset that are different from the SM. 
Moreover, when KMAD is used to constrain the operator coefficients, KMAD is also a supervised learning algorithm because one need to provide the SM dataset.

The difference between the two approaches can be neglected if it is assumed that the number of NP signals is small enough to affect the positions of the centroids.
Both approaches do not need to know what the NP models are being searched for, and therefore NP specific analysis is not needed. 
As will be shown in the next section, the KMAD can also been introduced as an aid the the event selection strategy.

We use the second approach in this paper.
When the second approach is used, the operator coefficients are completely irrelevant if only the distribution characteristics of the anomalous scores are studied.

\begin{figure}[htbp]
\begin{center}
\includegraphics[width=0.98\hsize]{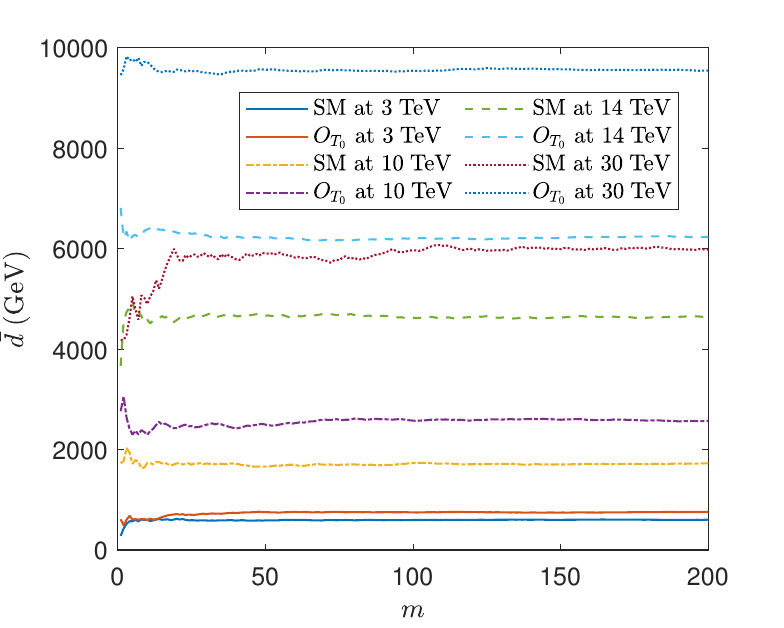}
\caption{\label{fig:history}$\bar{d}$ as functions of $m$ at $k=50$. We find that $\bar{d}$ converges rapidly as $m$ grows.}
\end{center}
\end{figure}

\begin{figure*}[htbp]
\begin{center}
\includegraphics[width=0.3\textwidth]{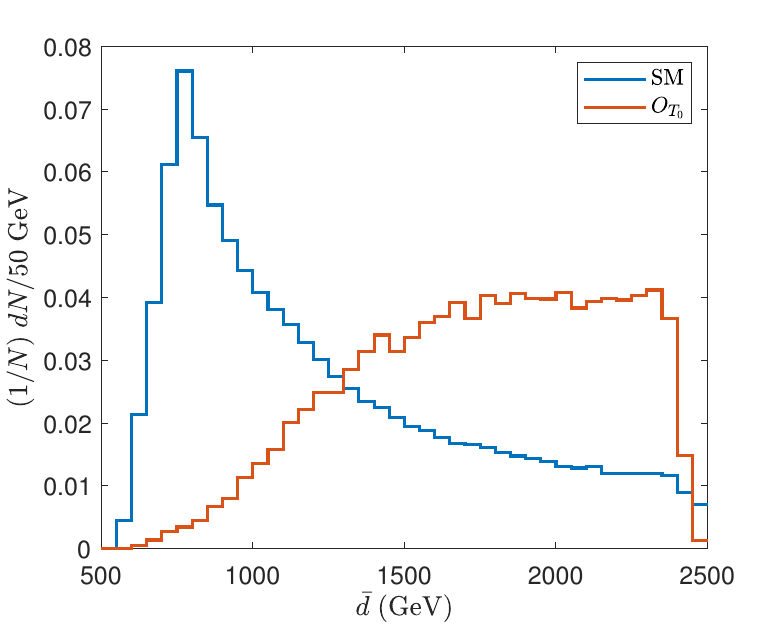}
\includegraphics[width=0.3\textwidth]{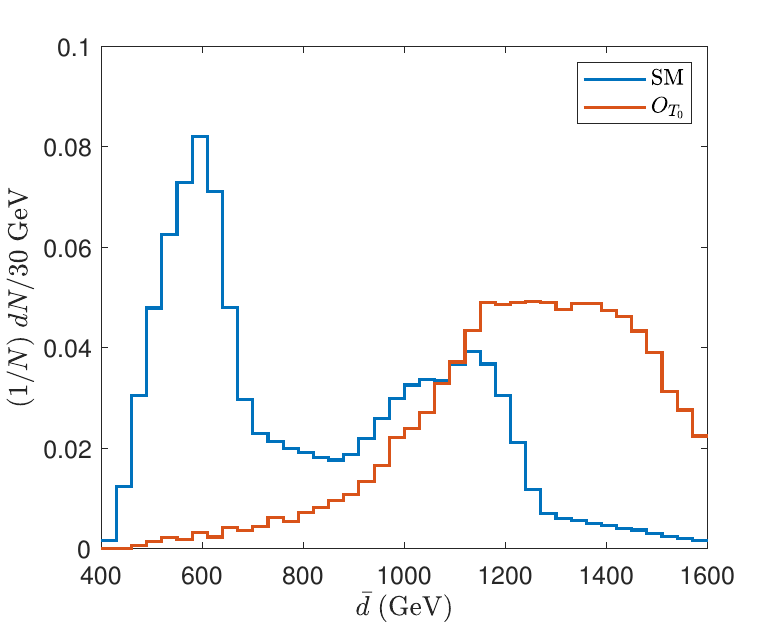}
\includegraphics[width=0.3\textwidth]{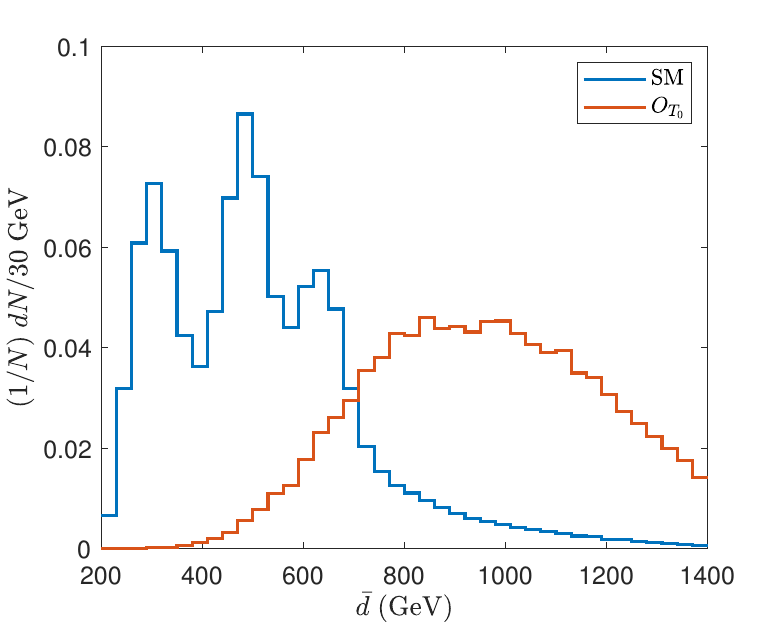}
\includegraphics[width=0.3\textwidth]{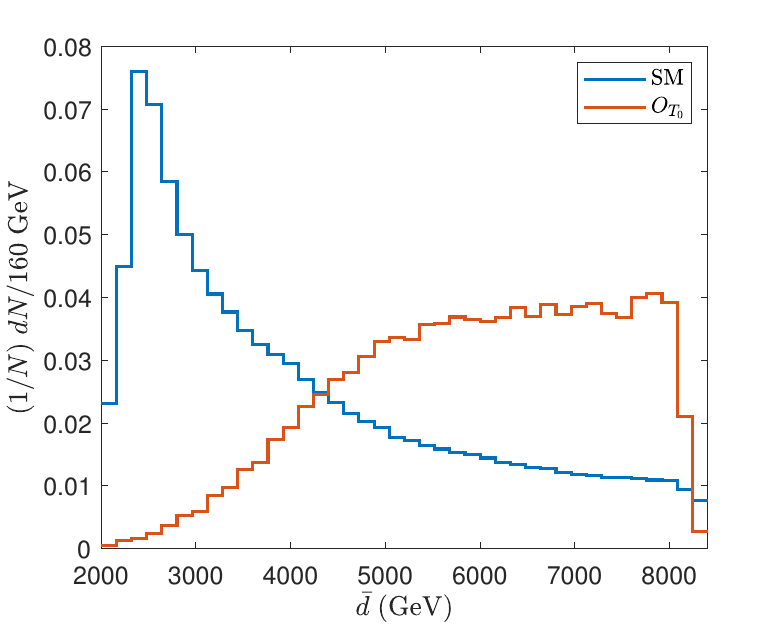}
\includegraphics[width=0.3\textwidth]{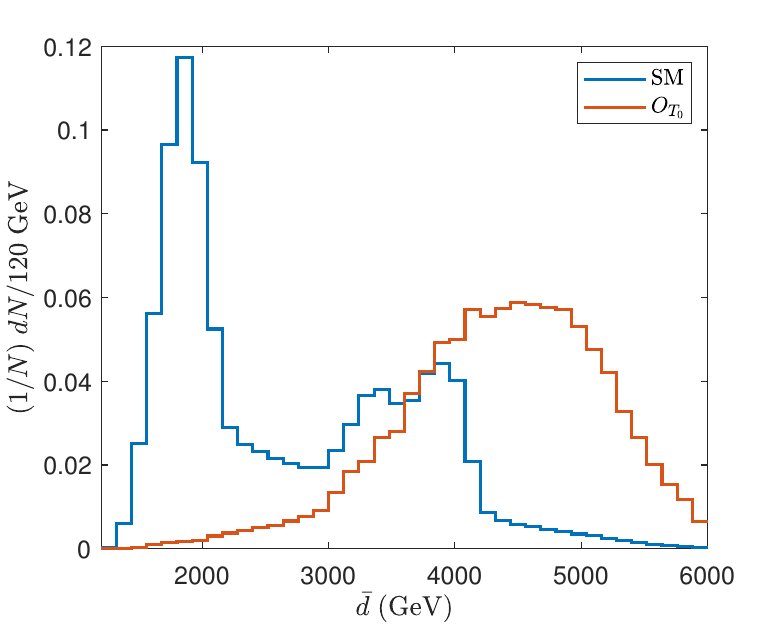}
\includegraphics[width=0.3\textwidth]{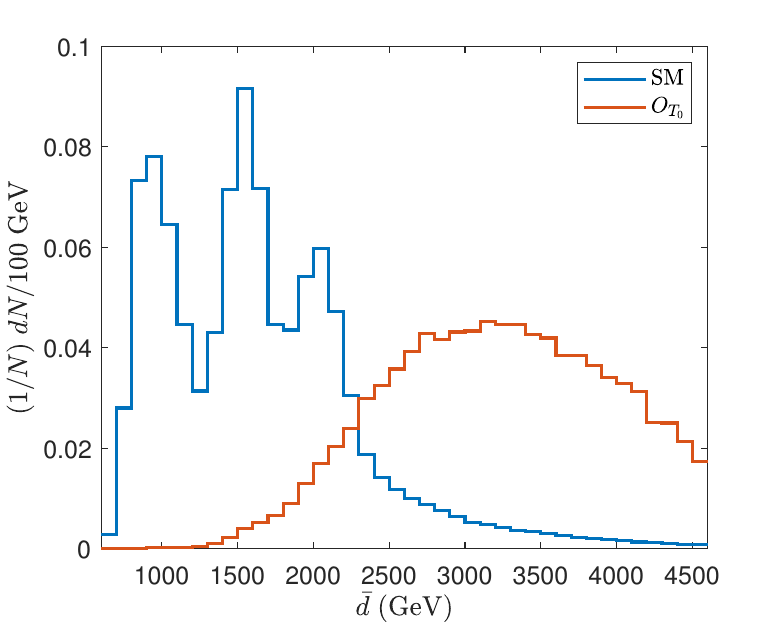}
\includegraphics[width=0.3\textwidth]{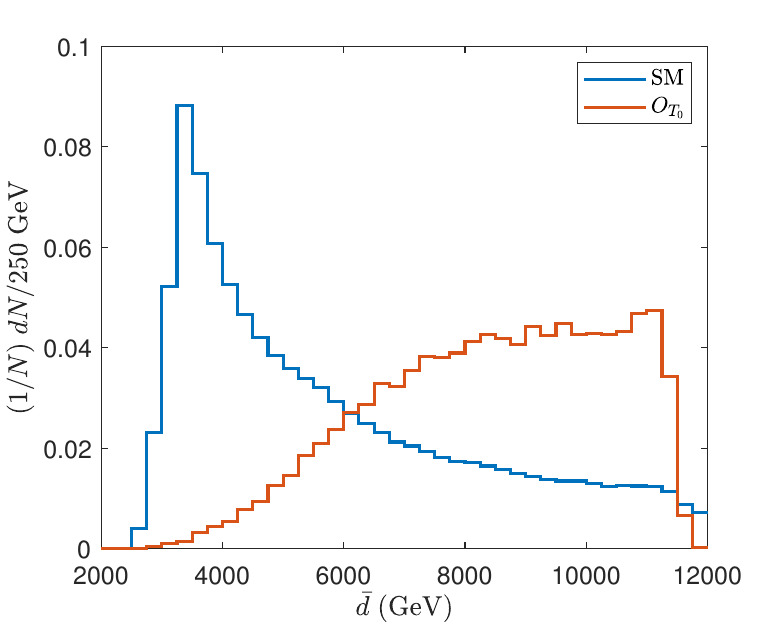}
\includegraphics[width=0.3\textwidth]{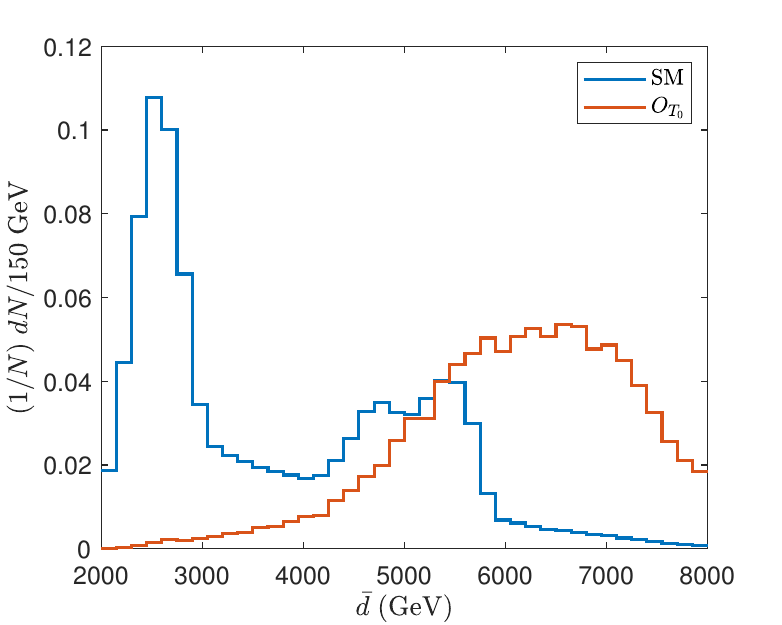}
\includegraphics[width=0.3\textwidth]{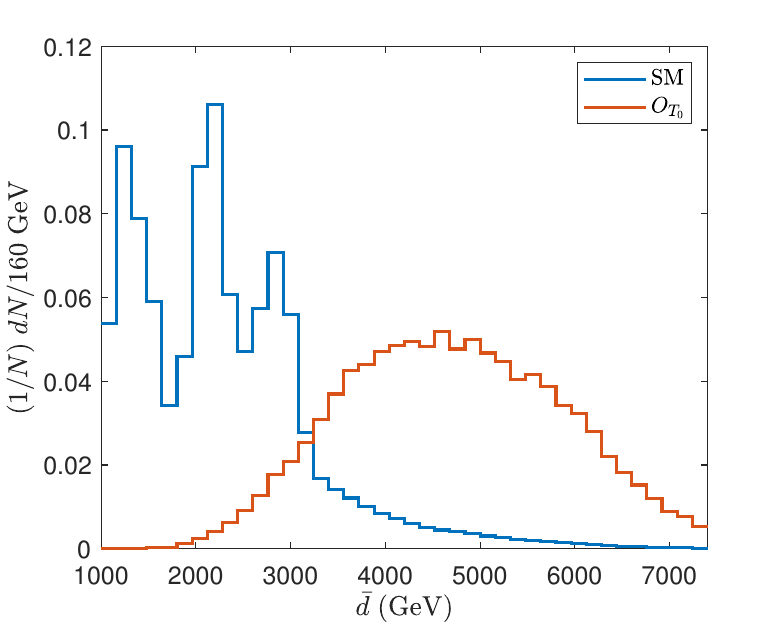}
\includegraphics[width=0.3\textwidth]{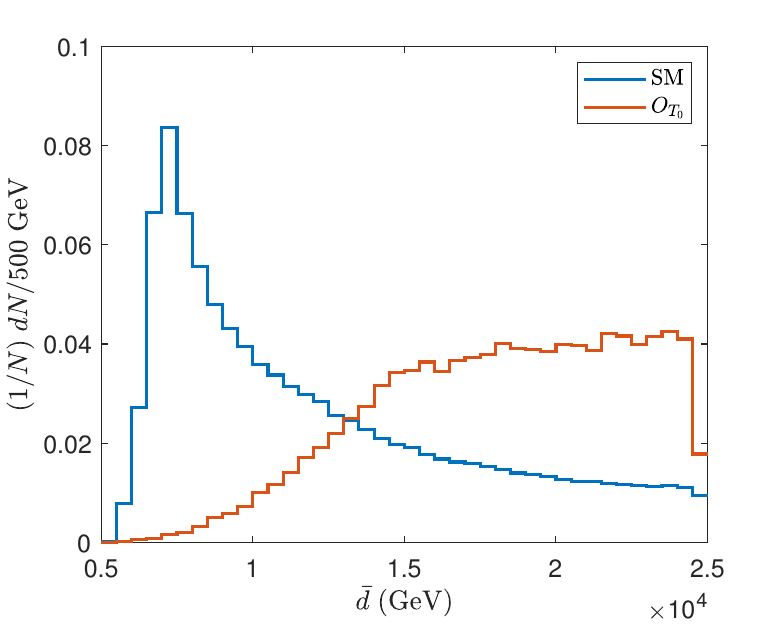}
\includegraphics[width=0.3\textwidth]{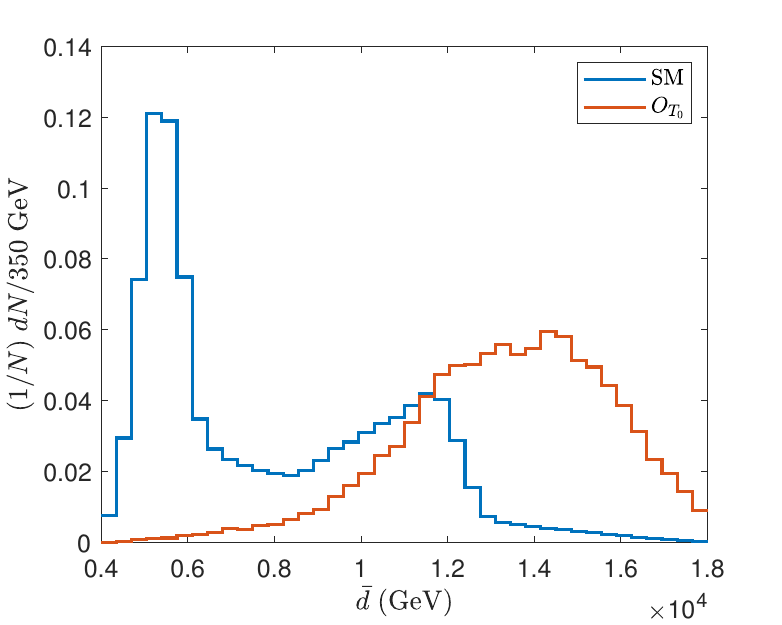}
\includegraphics[width=0.3\textwidth]{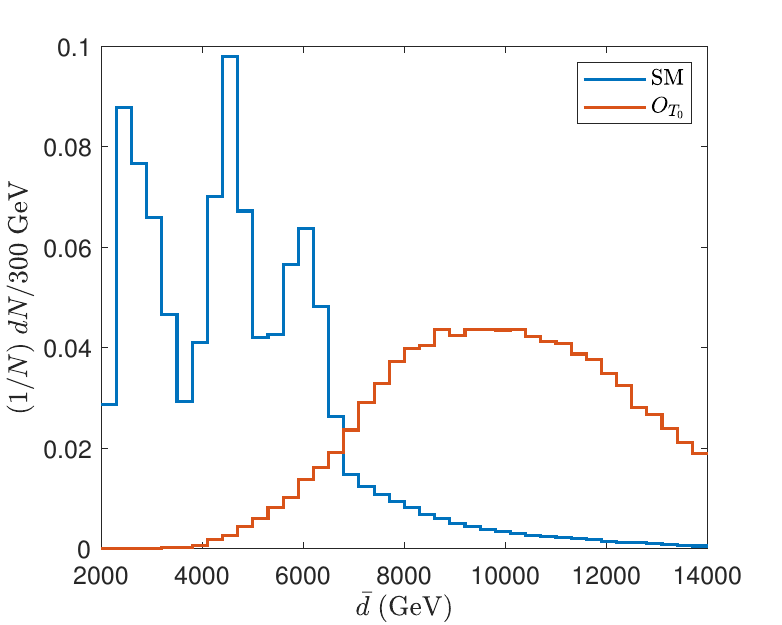}
\caption{\label{fig:15000K50}The normalized distribution of anomaly score $\bar{d}$ when $k=2$~(column 1), 10~(column 2), and 50~(column 3), at $\sqrt{s}=3\;{\rm TeV}$~(row 1), $10\;{\rm TeV}$~(row 2), $14\;{\rm TeV}$~(row 3) and $30\;{\rm TeV}$~(row 4) for the SM and $O_{T_0}$ induced contribution. }
\end{center}
\end{figure*}

As introduced, the number of repetitions of the KMAD process $m$ is a tunable parameter.
$m$ can be used to control the accuracy of the event selection strategy.
We studied the rate of convergence of $\bar{d}$ as $m$ grows.
Picking an event from the SM background and $O_{T_0}$ signals at 3\;{\rm TeV}, 10\;{\rm TeV}, 14\;{\rm TeV} and 30\;{\rm TeV}, $\bar{d}$ as functions of $m$ at $k=50$ are shown in Fig.~\ref{fig:history}.
We find that $\bar{d}$ converges rapidly as $m$ grows, when $m=200$, the value of $\bar{d}$ starts to become stable.
Theoretically, the value of $m$ can continue to increase to make the relative statistical error of $\bar{d}$ smaller, but due to limited computational power, we use $m=200$ in this paper.
The relative statistical error of $\bar{d}$ for each point does not exceed 4$\%$.

Another tunable parameter of KMAD is $k$.
If it is assumed that different processes are Gaussian distributed in momentum space, then from the point of view of computational efficiency, the best choice of $k$ should be the number of processes, i.e., each classification corresponds to a process of the background, and the NP is relatively farther away from the center of each Gaussian distribution. 
It is true that there is a possibility that each process in the SM corresponds to a Gaussian distribution in the space of observables by choosing the appropriate observables, however, given the interference in the different processes, the more general case is not encountered as a superposition of Gaussian distributions. 
Regardless of the case, a larger $k$ provides a better sample of the background distribution as shown in Fig.~\ref{fig:illustrative} while requiring more computational resources, so we choose a sufficiently large $k$ in a balance between accuracy and computational efficiency.
Fig.~\ref{fig:15000K50} shows the distribution of the SM and NP anomaly scores at different $k$.
It can be seen that the distributions of the anomaly scores for the SM background and the NP signals are different.
With limited computational power, we take the value of $k$ as $50$ in this paper.

\section{\label{sec4}Constraints on the coefficients}

\begin{table*}[htbp]
\centering
\begin{tabular}{c|c|c|c|c} 
\hline
 $\sqrt{s}$ & 3\;{\rm TeV} &10\;{\rm TeV} & 14\;{\rm TeV} & 30\;{\rm TeV} \\ 
 \hline
 Unit of coefficient & $ ({\rm TeV}^{-4})$ & $(10^{-3}\;{\rm TeV}^{-4})$ &$ (10^{-3}\;{\rm TeV}^{-4})$&$(10^{-4}\;{\rm TeV}^{-4})$\\
\hline
 $f_{T_0}/\Lambda ^4$ &[-0.3, 0.3]&[-1.5, 1.5] &[-0.5, 0.5]& [-0.5, 0.5]\\ 
\hline
 $f_{T_2}/\Lambda ^4$ & [-0.5, 0.5]&[-2.0, 2.0] & [-0.8, 0.8]& [-0.8, 0.8]\\
\hline
 $f_{T_5}/\Lambda ^4$ & [-0.06, 0.06] &[-0.3, 0.3]& [-0.1, 0.1]& [-0.08, 0.08]\\
\hline
 $f_{T_7}/\Lambda ^4$ & [-0.1, 0.1]& [-0.5, 0.5] &[-0.15, 0.15] &[-0.15, 0.15] \\
\hline
 $f_{T_8}/\Lambda ^4$ &[-0.01, 0.01]& [-0.05, 0.05]& [-0.015, 0.015]& [-0.015, 0.015] \\ [1ex]
\hline
 $f_{T_9}/\Lambda ^4$ &[-0.016, 0.016]& [-0.08, 0.08]& [-0.02, 0.02]& [-0.02, 0.02] \\ [1ex] 
\hline
\end{tabular}
\caption{The range of coefficients used in scanning.}
\label{table:2}
\end{table*}

When no NP signals are found, the study of NP is to set constraints on the parameters of NP.
The KMAD can also be used to set constraints on the coefficients of operators contributing to aQGCs.
For this purpose, we generate events by using MC with coefficients in Table.~\ref{table:2}.
The contribution of interference between the SM and aQGCs is also included.
For each value of the coefficients, 300000 events are generated.
It has been shown that $p_{T,\gamma}$ provides an efficient cut to suppress the SM background.
In Ref.~\cite{triphoton}, $p_{T,\gamma} > 0.12 E_{\rm beam}$ is used as a part of the event selection strategy where $E_{\rm beam}$ is the energy of the beam.
To leave some room for KMAD, when generating events, in the standard cuts $p_{T,\gamma} > 0.1 E_{\rm beam}$ is used, other standard cuts relevant to infrared divergences are as same as those in Eq.~(\ref{eq.standardcuts}).
The beam induced background is also very important, for example, the radiated photons tangent to the muon trajectories. 
At this stage, it is difficult to consider the impact of this part of the contribution since Delphes does not yet include it. 
However, due to the inclusion of the $p_T$ cut in the standard cuts, the impact of this part will be significantly removed. 
The detailed quantitative analysis is deferred to future works.

Similar as the previous section, at least $3$ photons are required to present after fast detector simulation.
Then the KMAD event selection strategy is applied to select the events with $\bar{d}$ larger than $700\;{\rm GeV}$, $2000\;{\rm GeV}$, $3000\;{\rm GeV}$ and $7000\;{\rm GeV}$ for $\sqrt{s} = 3\;{\rm TeV}$, $10\;{\rm TeV}$, $14\;{\rm TeV}$ and $30\;{\rm TeV}$, respectively.
In this paper, the kinematic cuts are used, so this work can only be regarded as using the AD algorithm to improve the efficiency of the event selection strategy, and the event selection strategy in this paper is a combination of traditional and machine learning strategies.
The standard cuts used to generate events are used to avoid infrared divergence, and the required number of end-state photons depends on the process to be studied, none of which requires the knowledge of the NP model being studied.
Of course those cuts inevitably affect the efficiency of the event selection strategy, especially that we use a large $p_{T,\gamma}$ cut to suppress the SM background.

\begin{figure*}[htpb]
\begin{center}
\includegraphics[width=0.3\textwidth]{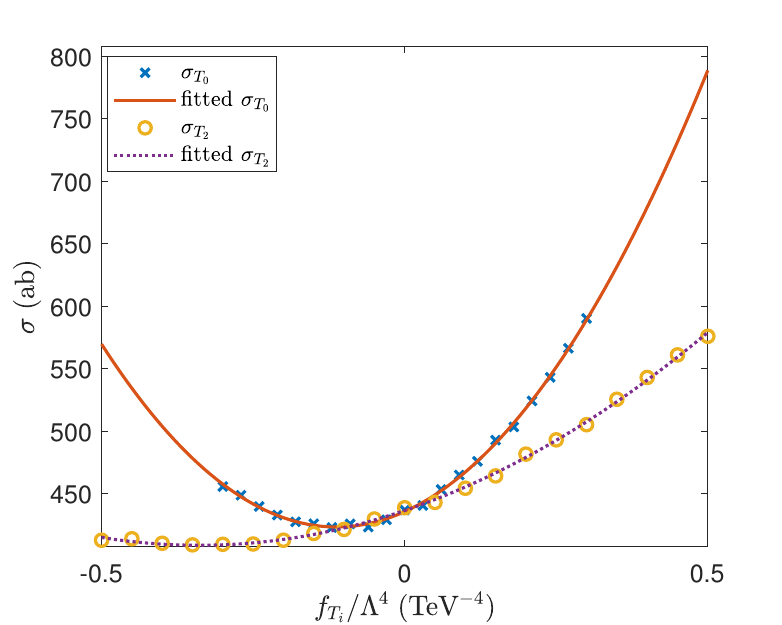}
\includegraphics[width=0.3\textwidth]{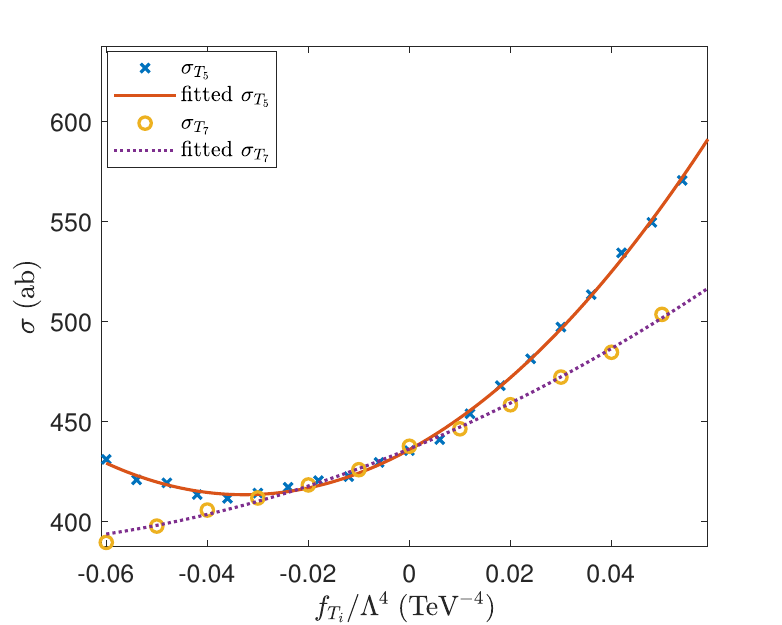}
\includegraphics[width=0.3\textwidth]{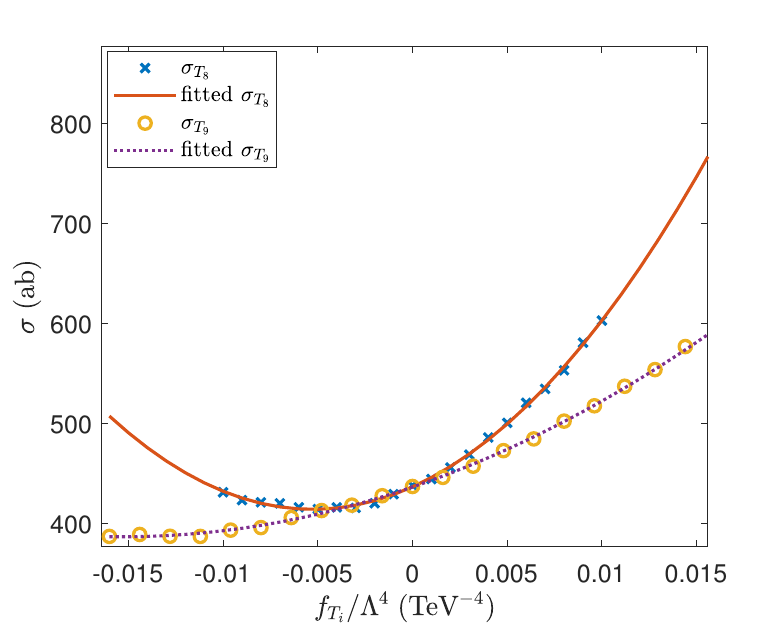}
\includegraphics[width=0.3\textwidth]{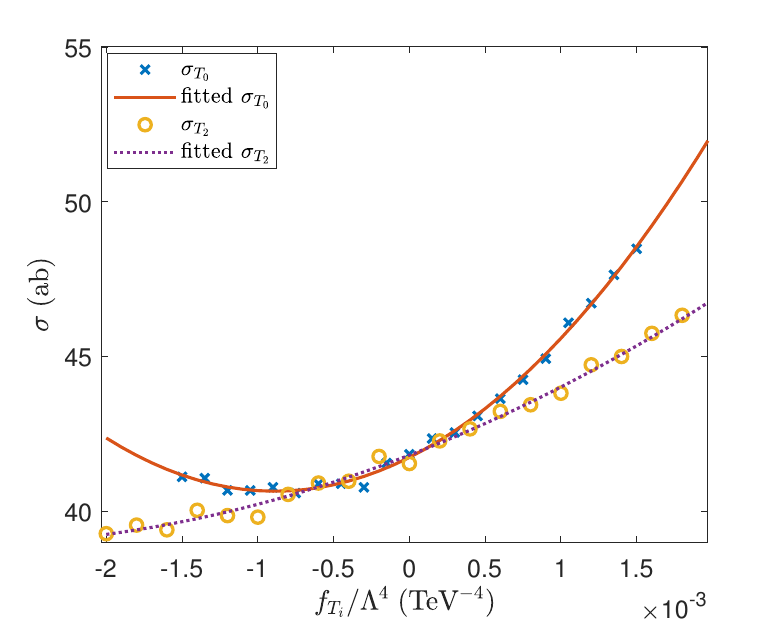}
\includegraphics[width=0.3\textwidth]{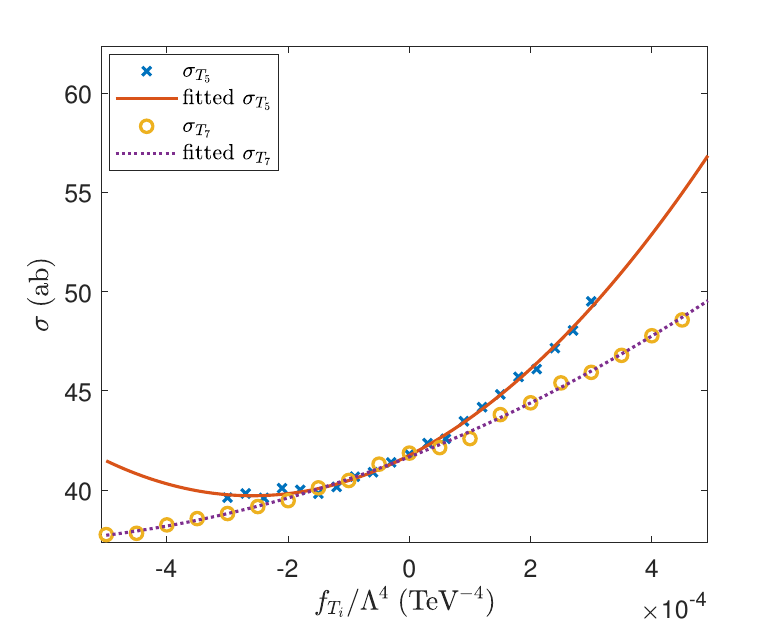}
\includegraphics[width=0.3\textwidth]{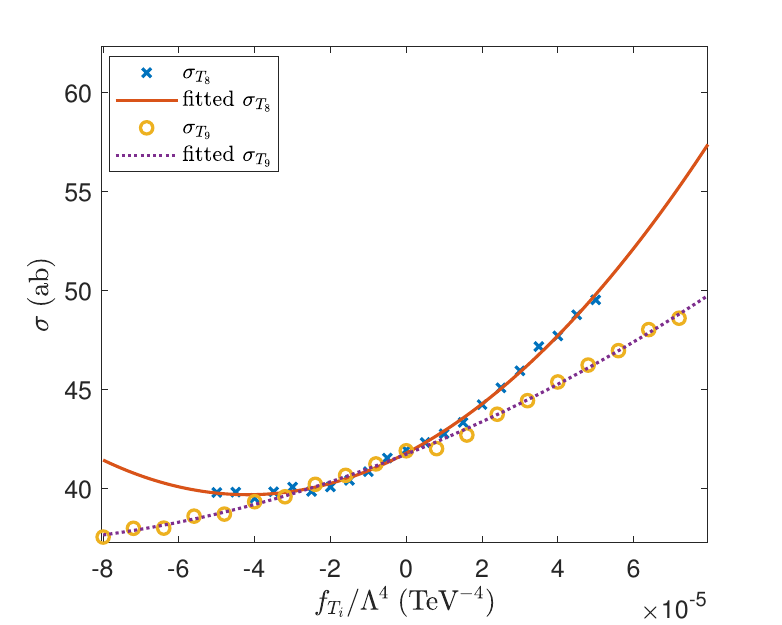}
\includegraphics[width=0.3\textwidth]{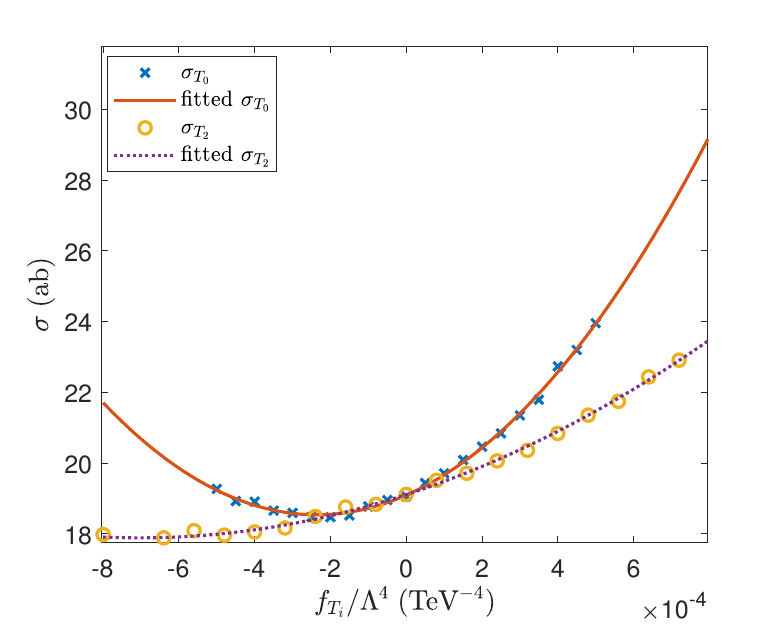}
\includegraphics[width=0.3\textwidth]{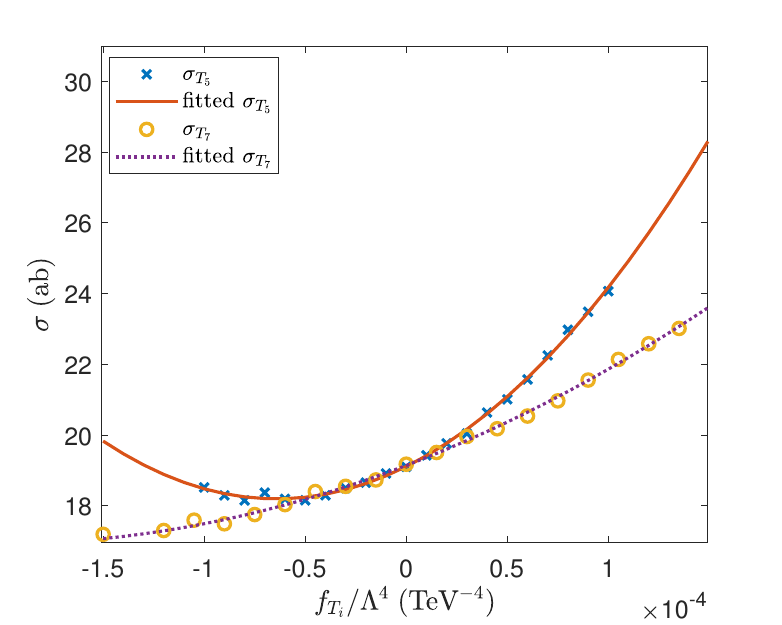}
\includegraphics[width=0.3\textwidth]{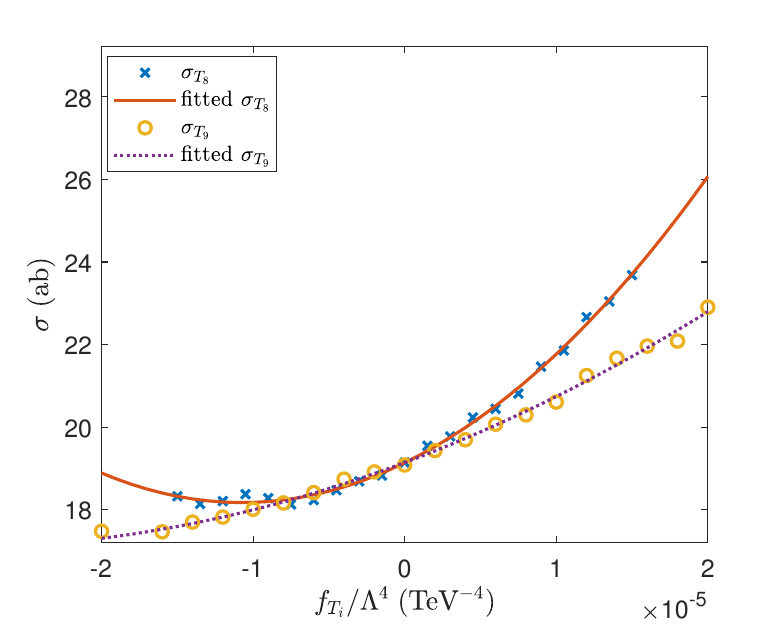}
\includegraphics[width=0.3\textwidth]{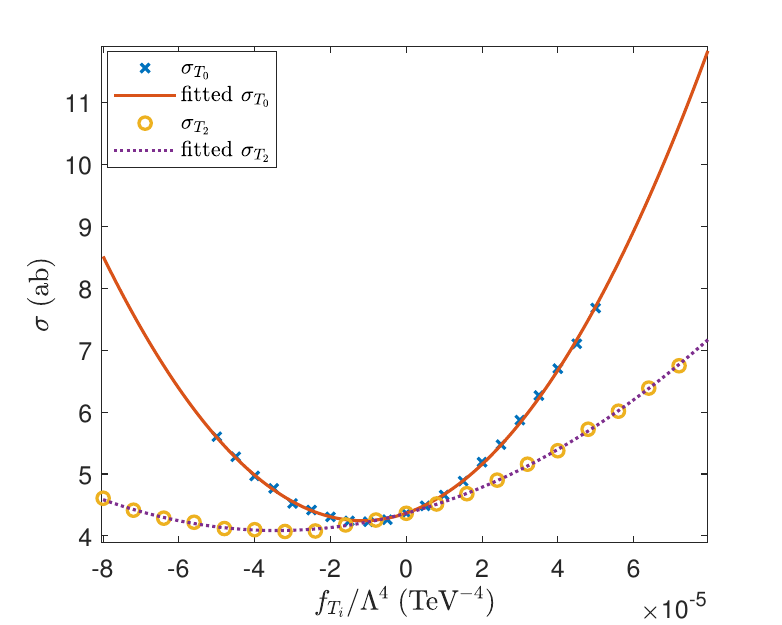}
\includegraphics[width=0.3\textwidth]{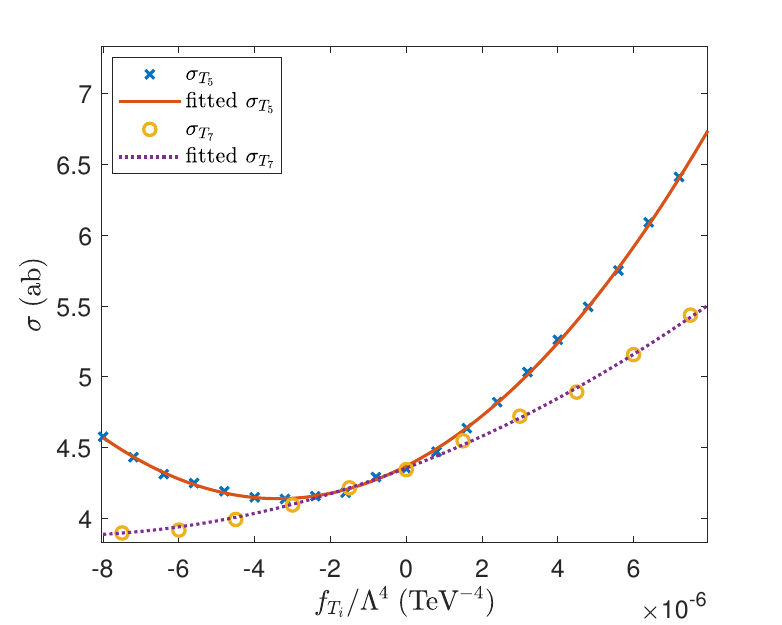}
\includegraphics[width=0.3\textwidth]{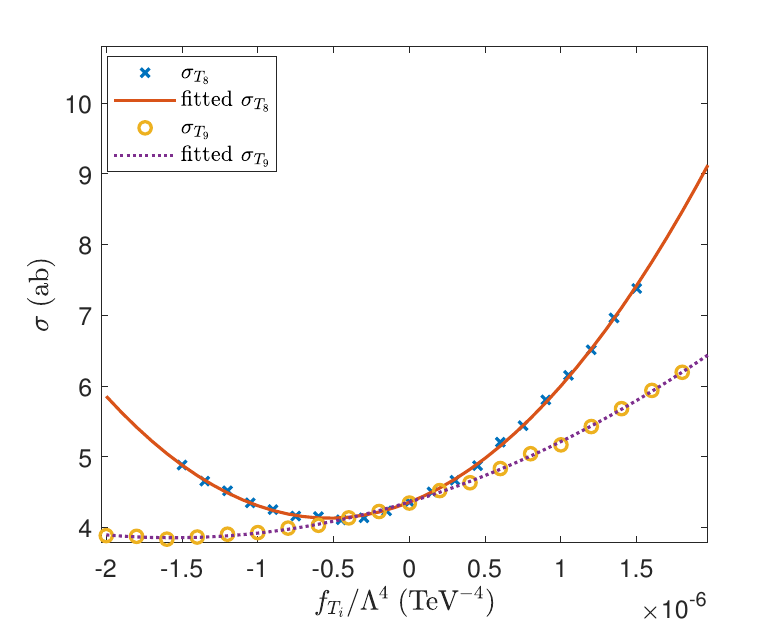}
\caption{\label{fig:15000t8t9}
The cross-sections $\sigma$ and fitted $\sigma$ as bilinear functions of $f_{T_{i}}/\Lambda^4$ at $\sqrt{s}$ = 3\;{\rm TeV} (row 1), 10\;{\rm TeV} (row 2), 14\;{\rm TeV}  (row 3) and 30\;{\rm TeV}  (row 4). }
\end{center}
\end{figure*}

With aQGCs presented, the cross-section of tri-photon process can be written as\begin{equation}
\begin{split}
&\sigma=\sigma_{\rm SM}+ \sigma_{\rm int} f_{T_{i}}/\Lambda^{4}+ \sigma_{\rm NP} \left(f_{T_{i}}/\Lambda^{4}\right)^2,
\end{split}
\label{eq.fit}
\end{equation}
where $\sigma _{\rm SM}$ is the SM contribution, $\sigma _{\rm int} f_{T_{i}}/\Lambda^{4} $ denotes interference between the SM and aQGCs, $\sigma _{\rm NP} \left(f_{T_{i}}/\Lambda^{4}\right)^2$ is the aQGCs induced contribution.
The analytical results of $\sigma _{\rm int}$ and $\sigma _{\rm NP}$ before cuts can be found in Ref.~\cite{triphoton}.
After event selection strategy, the cross-section is fitted as a bilinear function of $f_{T_{i}}/\Lambda^{4}$ in Eq.~(\ref{eq.fit}).
The cross-sections after event selection strategy and the fitted cross-sections are shown in Fig.~\ref{fig:15000t8t9}.
It can be found from Fig.~\ref{fig:15000t8t9} that after the event selection strategy, the cross-sections still fit bilinear functions well when the coefficients are within the range listed in Table~\ref{table:2}.
The interference between the SM and aQGCs plays an important role in the tri-photon process.
From Fig.~\ref{fig:15000t8t9}, it can be found that the KMAD works well when the interference presents.

\begin{table*}[htbp]
\centering
\begin{tabular}{c|c|c|c|c|c} 
\hline
  &  &3\;{\rm TeV} &10\;{\rm TeV} &14\;{\rm TeV} &30\;{\rm TeV} \\
  &$S_{stat}$& $1\;{\rm ab}^{-1}$&$10\;{\rm ab}^{-1}$&$10\;{\rm ab}^{-1}$ &$10\;{\rm ab}^{-1}$ \\
  &  &$(10^{-2}\;{\rm TeV^{-4}})$&$(10^{-4}\;{\rm TeV^{-4}})$&$(10^{-4}\;{\rm TeV^{-4}})$&$(10^{-5}\;{\rm TeV^{-4}})$ \\
\hline
  & 2 &[-34.94, 12.46]&[-28.31,10.57]& [-8.19, 3.47]&[-5.23, 2.91]\\
$f_{T_{0}}(f_{T_{1}})/\Lambda^4$& 3 &[-39.25, 16.77]&[-31.92, 14.48]& [-9.34, 4.61]&[-6.13, 3.81] \\
  & 5 & [-46.45, 23.97]&[-37.93, 20.19]&[-11.25, 6.53]&[-7.63, 5.31]\\
\hline
 & 2 &[-33.92, 19.99]&[-78.77, 17.16]&[-19.83, 5.77]&[-11.79, 4.95] \\
$f_{T_{2}}/\Lambda^4$& 3 &[-95.61, 27.76]&[-85.51, 23.90]&[-21.96, 7.90]&[-13.48, 6.64] \\
  & 5 &[-109.03, 41.19]&[-97.20, 35.58]&[-25.60, 11.55]&[-16.35, 9.51] \\
\hline
 & 2 &[-8.81, 2.28]&[-7.11, 1.92]&[-1.94, 0.658]& [-1.23, 0.557]\\
$f_{T_{5}}(f_{T_{6}})/\Lambda^4$& 3 &[-9.67, 3.14]&[-7.82, 2.64]&[-2.17, 0.891]&[-1.42, 0.742] \\
  & 5 &[-11.14, 4.61]&[-9.05, 3.86]&[-2.57, 1.28]&[-1.73, 1.06] \\
\hline
& 2 &[-13.37, 3.46]&[-5.17, 2.89]&[-5.10,1.03]& [-2.80, 0.917]\\
$f_{T_{7}}/\Lambda^4$& 3 &[-23.92, 4.92]&[-7.34, 4.09]&[-5.52, 1.45]&[-3.14, 1.26] \\
  & 5 &[-26.51, 7.51]&[-20.91, 6.22]&[-6.25, 2.18]& [-3.73, 1.84] \\
\hline
 &2 &[-1.42, 0.368]&[-1.14, 0.304]&[-0.321,  0.106]&[-0.195, 0.0890] \\
$f_{T_{8}}/\Lambda^4$& 3 &[-1.56, 0.507]&[-1.25, 0.418]&[-0.358, 0.144]&[-0.225, 0.119] \\
  & 5 &[-1.80, 0.745]&[-1.45, 0.612]&[-0.422,  0.207]& [-0.275, 0.169]\\
\hline
 &2 &[-0.895, 0.554]&[-2.90, 0.461]&[-0.767, 0.163]&[-0.458, 0.146] \\
$f_{T_{9}}/\Lambda^4$& 3 &[-1.53, 0.787]&[-3.09, 0.653]&[-0.832, 0.227]& [-0.513, 0.200]\\
  & 5 &[-4.27, 1.20]&[-3.44, 0.995]&[-0.945, 0.341]&[-0.607, 0.294]\\
\hline
\end{tabular}
\caption{The projected sensitivities on the coefficients of the $O_{T_i}$ operators at the muon colliders with different c.m. energies and integrated luminosities for the ``conservative'' case.}
\label{table:3}
\end{table*}

\begin{table}[htbp]
\centering
\begin{tabular}{c|c|c|c} 
\hline
  &  &14\;{\rm TeV} &30\;{\rm TeV} \\
  &$S_{stat}$& $20\;{\rm ab}^{-1}$&$90\;{\rm ab}^{-1}$ \\
  &  &$(10^{-5}\;{\rm TeV^{-4}})$&$(10^{-6}\;{\rm TeV^{-4}})$ \\
\hline
  & 2 &[-74.12, 26.90]&[-36.61, 13.41] \\
$f_{T_{0}}(f_{T_{1}})/\Lambda^4$& 3 &[-83.39, 36.16]&[-41.21, 18.02] \\
  & 5 & [-98.88, 51.65]& [-48.90, 25.71] \\
\hline
 & 2 &[-184.18, 43.61]&[-89.47, 21.05] \\
$f_{T_{2}}/\Lambda^4$& 3 &[-201.00, 60.43]&[-97.60, 29.18] \\
  & 5 &[-230.00, 89.43]& [-111.62, 43.20] \\
\hline
 & 2 &[-17.86, 5.02]&[-9.17, 2.42] \\
$f_{T_{5}}(f_{T_{6}})/\Lambda^4$& 3 &[-19.72, 6.88]&[-10.08, 3.32] \\
  & 5 & [-22.88, 10.04]&[-11.63, 4.88] \\
\hline
& 2 &[-12.88, 7.66]&[-12.39, 3.68] \\
$f_{T_{7}}/\Lambda^4$& 3 &[-20.35, 10.87]&[-24.09, 5.022] \\
  & 5 &[-57.28, 16.58]& [-26.81, 7.93] \\
\hline
 &2 &[-2.95, 0.806]&[-1.45, 0.387] \\
$f_{T_{8}}/\Lambda^4$& 3 &[-3.25, 1.10]&[-1.59, 0.532] \\
  & 5 &[-3.77, 1.62]& [-1.84, 0.780] \\
\hline
 &2 &[-2.21, 1.21]&[-0.972, 0.583] \\
$f_{T_{9}}/\Lambda^4$& 3 &[-7.75, 1.71]&[-1.56, 0.828] \\
  & 5 &[-8.64, 2.60]&[-4.39, 1.26] \\
\hline
\end{tabular}
\caption{Same as Table~\ref{table:3} but for the ``optimistic'' case.}
\label{table:4}
\end{table}

The constraints on the coefficients are estimated by using statistical sensitivity defined as~\cite{Cowan:2010js,pdg}
\begin{equation}
\begin{split}
&\mathcal{S}_{stat}=\sqrt{2 \left[(N_{\rm bg}+N_{s}) \ln (1+N_{s}/N_{\rm bg})-N_{s}\right]},
\end{split}
\end{equation}
where $N_s=(\sigma-\sigma_{\rm SM})L$ and $N_{\rm bg}=\sigma_{\rm SM}L$, and $L$ is the luminosity.
In this paper, the luminosities in both the ``conservative'' and ``optimistic'' cases~\cite{muoncollider5} are considered.
The constraints on the coefficients at $S_{stat}= 2, 3, 5$ are listed in Tables~\ref{table:3} and \ref{table:4}.

\begin{figure*}[htpb]
\begin{center}
\includegraphics[width=0.24\textwidth]{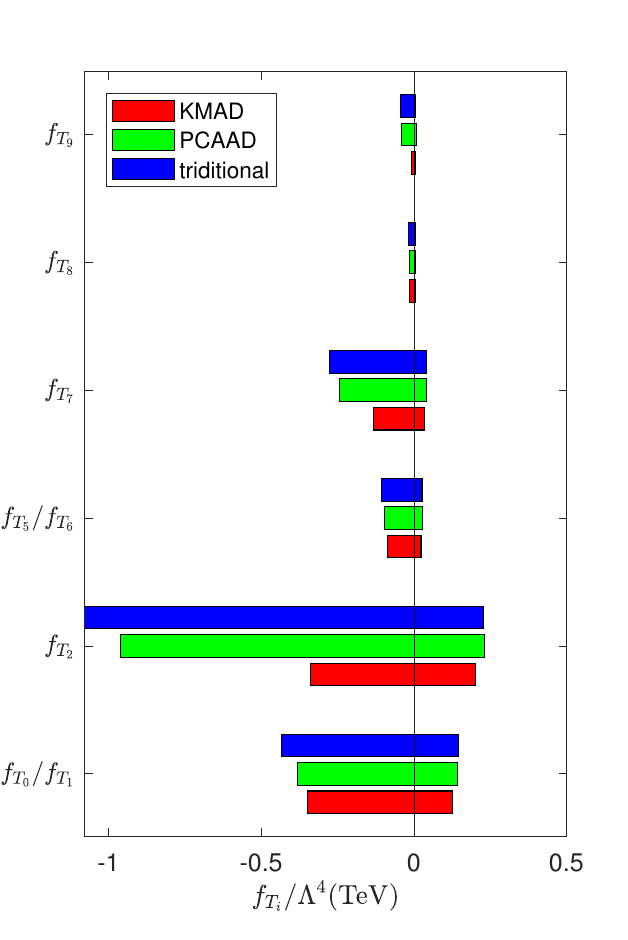}
\includegraphics[width=0.24\textwidth]{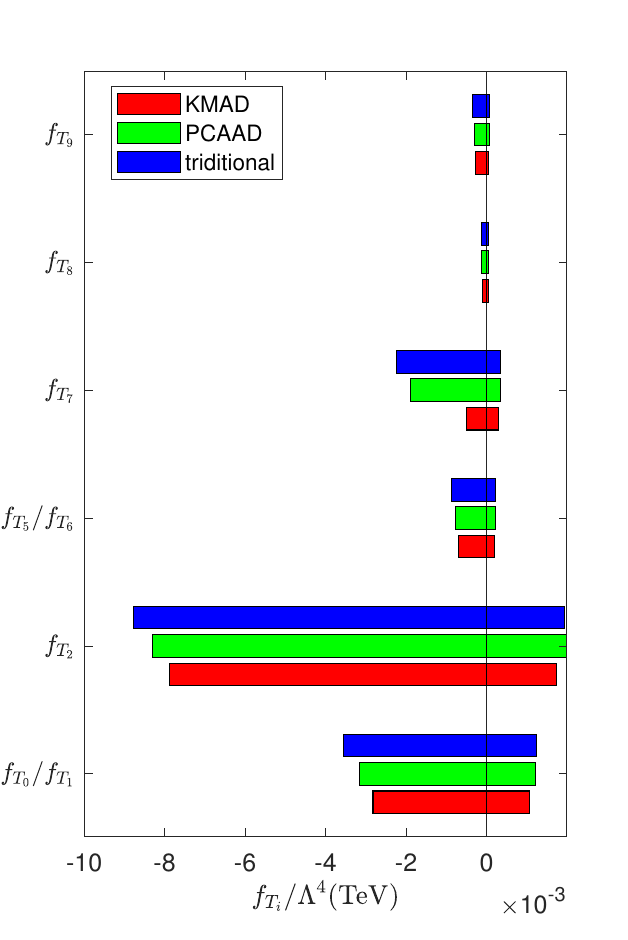}
\includegraphics[width=0.24\textwidth]{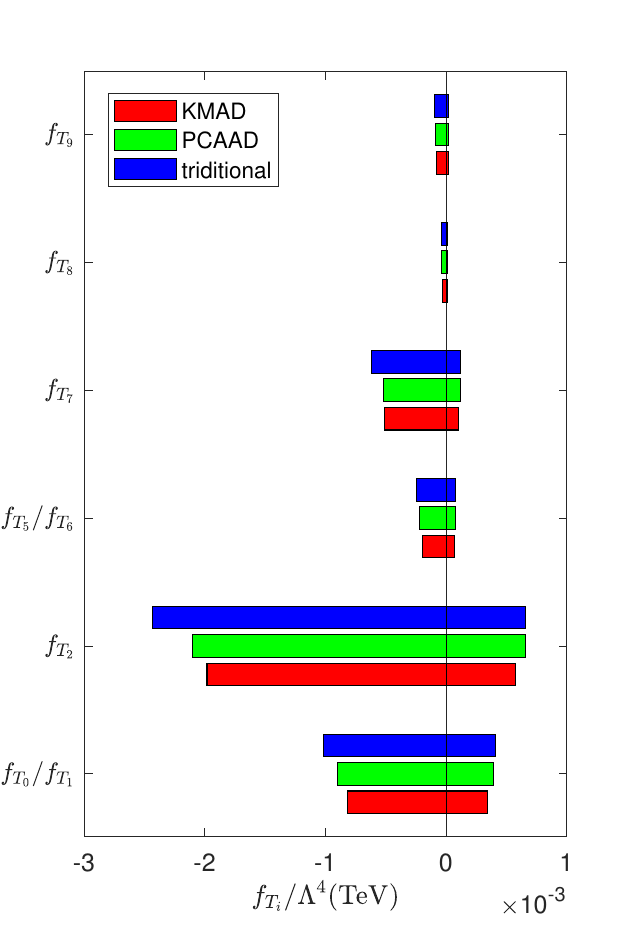}
\includegraphics[width=0.24\textwidth]{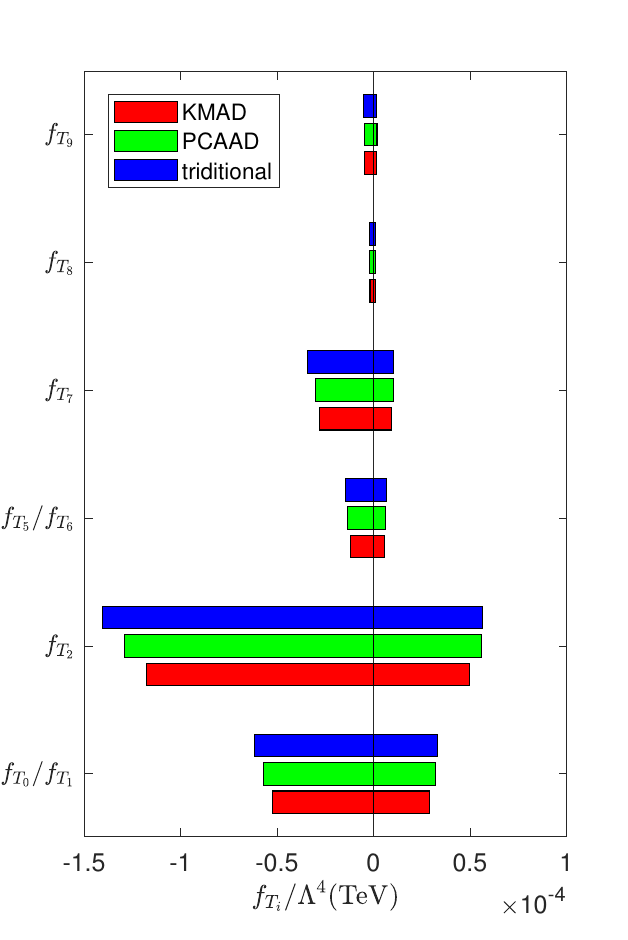}
\caption{\label{fig:compare}
The expected constraints at $95\%$ C.L. in the ``conservative'' case using different event selection strategies. The panels from left to right correspond to $\sqrt{s}=3,10,14$ and $30$ TeV, respectively. The ``KMAD'', ``PCAAD'' and ``traditional'' correspond to the strategy in this paper, in Ref.~\cite{Dong:2023nir} and in Ref.~\cite{triphoton}, respectively.}
\end{center}
\end{figure*}
In Refs.~\cite{triphoton,Dong:2023nir}, the same process and same operators are studied.
The expected constraints at $95\%$ C.L. in the ``conservative'' case are compared in Fig.~\ref{fig:compare}, where ``KMAD'', ``PCAAD'' and ``traditional'' correspond to the strategy in this paper, in Ref.~\cite{Dong:2023nir} and in Ref.~\cite{triphoton}, respectively.
It can be seen from Fig.~\ref{fig:compare} and Tables~\ref{table:3} and \ref{table:4} that, the expected constraints in this paper are generally strengthened, some are even about one order of magnitude tighter than those in Ref.~\cite{triphoton}.
Compared with Ref.~\cite{Dong:2023nir} which also concentrate on a machine learning algorithm which has the potential to be accelerated by quantum computers, the KMAD is also better.
It can be concluded that, the KMAD is useful and efficient in the search of NP signals.
There may be other observables for this process that can be used to distinguish the NP signal from the background. 
However, unlike the KMAD, the effectiveness of different observables often depends on what the NP to be sought is. 
The comparisons with traditional event selection strategy in this paper already show this advantage, and for simplicity we neglect to discuss other traditional event selection strategies.

Note that different datasets are used in this section which also verifies that overfitting is not causing a significant impact.
Overfitting means that the algorithm is finding patterns only exist in the training dataset, and don't generalize to new, unseen data.
In training, the KMAD trains the centroids, which are obtained using the SM dataset.
When constraining the operator coefficients, the events are generated with the SM, NP and interference contributions involved.
In experiments, the dataset used to constrain the operator coefficients should come from measurements.
Therefore, the results in this section~(or in experiments) use datasets different from the one in training, which can be regarded as new and unseen data during the training phase.  

\section{\label{sec5}Summary}

Searching for NP signals requires processing a large amount of data, while it has been shown that the k-means algorithm is able to be accelerated by using quantum computers, which are capable of handling large amounts of data, so it is a question worth investigating whether the k-means algorithm can also be used to search for NP signals.
In this paper, taking the case of aQGCs in the tri-photon processes at the muon colliders as an example, we propose an event selection strategy using KMAD which is based on the k-means algorithm to search for NP signals.

By using MC, the expected constraints on the coefficients of the operators contributing to aQGCs are calculated.
The upper bounds in this paper are generally tighter than those obtained by using a traditional event selection strategy.
It can be concluded that, the KMAD is useful and efficient in the search of NP signals.

The ability of quantum computing to handle large amounts of data will provide a whole new opportunity in the study of NP. 
Since the k-means algorithm can be accelerated by quantum computing and, at the same time, the k-means algorithm can be used to search for NP signals, we expect that the quantum computers can be used to help to search for NP in the near future.

\begin{acknowledgement}
This work was supported in part by the National Natural Science Foundation of China under Grants Nos. 11905093 and 12147214, the Natural Science Foundation of the Liaoning Scientific Committee No.~LJKZ0978 and the Outstanding Research Cultivation Program of Liaoning Normal University (No.21GDL004).
\end{acknowledgement}

\bibliography{kmeans}
\bibliographystyle{elsarticle-num}  

\end{document}